\documentstyle[prl,aps,multicol,epsf]{revtex}

\input{psfig.sty}

\renewcommand{\narrowtext}{\begin{multicols}{2}
\global\columnwidth20.5pc} 
\renewcommand{\widetext}{\end{multicols}
\global\columnwidth42.5pc} \multicolsep = 8pt plus 4pt minus 3pt

\begin{document}
\draft
\title{Linewidth of a polariton laser: Theoretical analysis of self-interaction
effects}
\date{\today}
\author{D. Porras and C. Tejedor}
\address{Departamento de F\'{\i}sica Te\'{o}rica de la Materia Condensada. 
Universidad Auton\'{o}ma de Madrid. 28049 Cantoblanco, Madrid, Spain.}
\maketitle

\begin{abstract}
Polaritons in semiconductor microcavities can experience a Bose-Einstein 
condensation experimentally detectable in the spectrum of the emitted light. 
Scattering with noncondensed particles as well as self-interaction in 
the condensate provoke phase-diffusion limiting the coherence of the polariton 
condensate. 
We present a theoretical analysis of self-interaction 
effects on the lineshape of the emission from a polariton laser.
Our calculations, for CdTe microcavities, show that there is an optimum 
pump at which the linewidth of the emitted light is reduced down to $1 \mu eV$.  
\end{abstract}


\narrowtext

A Bose-Einstein atomic condensate is a source of coherent matter-waves (matter 
laser), in the same way as a photonic laser is a source of coherent light.
Another system capable of undergoing a similar condensation is that of
polaritons formed by the strong coupling between quantum well excitons
and confined photons in a semiconductor microcavity. Polaritons 
behave as composite bosons at densities below the saturation density, 
as confirmed by recent experiments
that include the observation of stimulated scattering, and parametric
amplification and oscillation \cite{parametric,Ciuti}.
A polariton Bose-Einstein condensate 
is a matter laser that can be optically pumped, and experimentally detected by the
emitted light \cite{Imamoglu,Eastham,Deng}.
The growth of semiconductor microcavities with new materials,
such as II-VI compounds, or GaN, opens great possibilities, due to the strong
stability of the exciton in these systems \cite{CdTe,Zamfirescu}.
In particular, a recent calculation \cite{Porras}
shows that in high quality CdTe microcavities, huge occupation numbers 
can be achieved in the polariton ground state at densities well below the
saturation density, i. e., at densities at which polaritons can be described
as interacting bosons. 
Under these conditions, the system would be unstable to symmetry-breaking, and thus,
to the formation of a BEC.

In this work, we establish experimental signatures of a polariton laser.
We consider a microcavity pumped by a continuous nonresonant
laser, so that, the polariton-polariton scattering is fast
enough to overcome the radiative losses and the system
is able to relax to the ground state. 
At densities larger than a given threshold for BEC, the lowest energy level shows a
macroscopic occupancy, and the system becomes a continuous polariton 
laser. The standard theory of photon lasers \cite{Scully}, i. e. non-interacting 
systems, predicts a very narrow linewidth, inversely proportional to the number 
of condensed particles, $\Gamma^{NI} \propto 1/N_0$. 
However, polaritons interact with each other through some potential $V$ and 
new physics is involved. 
Self-interaction in the condensate provokes a process of
phase-diffusion that is determined by the energy scale $V N_0$
\cite{Holland,GardinerIII,Tassone00}. When $V N_0$ is comparable to 
$\hbar \Gamma^{NI}$, this process would increase the linewidth as the number of
condensed bosons increases, a behavior that is the opposite to that of a photon
laser. We introduce here a self-consistent framework to include the two effects
described above. We show that there is an optimum pumping range to get 
an extremely narrow linewidth of the polariton emission. For CdTe microcavities,
we find that the linewidth can be reduced down to $1 \mu eV$.
\begin{figure}
\centerline{
\psfig{figure=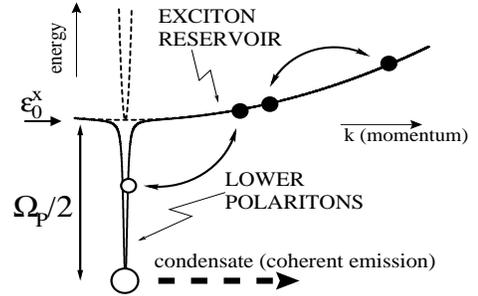,height=2.0in,width=2.5in}
}
\caption{Phase space for the scattering of polaritons in a semiconductor 
microcavity. Upper and lower
dashed lines represent the bare photon and bare exciton, respectively.
The continuous line is the polariton dispersion that
results from the strong exciton-photon coupling.}
\label{inout}
\end{figure}

Fig.\ref{inout} depicts schematically the lower energy branch of the spectrum
of a semiconductor microcavity for the case of zero detuning \cite{Hopfield}.
Below the minimum bare exciton energy, $\epsilon^x_0$,
excitons and photons are strongly coupled. 
The ground state 
($k=0$) lies at the energy $\Omega_P/2$ below $\epsilon^x_0$, 
where $\Omega_P$ is the polariton splitting.
At energies above $\epsilon^x_0$, 
the polariton branch merges abruptly into the bare exciton dispersion, and the
exciton-photon coupling can be neglected. We consider the phase space as
divided in two regions: the lower energy states, labelled as lower 
polaritons (LP), and the states above $\epsilon^x_0$, treated as bare excitons.
The small density of states of the LP, when compared to the
exciton one ($\rho_{LP} \approx 10^{-4} \rho_x$), drastically reduces the
threshold density for the formation of a BEC in the LP part of the spectrum.  
This allows us to simplify the problem of the polariton dynamics by
considering the exciton states as a thermalized reservoir\cite{Porras}.
On the other hand,
microcavity polaritons are 2-D quasiparticles
that can undergo a Kosterlitz-Thouless transition in the thermodynamic limt
\cite{Malpuech2}. Finite size effects leads to a local transition to BEC, so that we
quantize the LP levels according to a scale determined by area $S$ of the 
spot of the pumping laser.


In this work we are not analyzing the true quantum state of the polariton
condensate, but we are interested in the spectrum of the emitted light, that can be
calculated from the adequate correlation function.
The steps of the theoretical analysis are: first, to obtain 
an equation of motion for the density matrix, and later to compute correlation 
functions and emitted intensities.

{\it Equation of motion for the density matrix.}
The main mechanism for the relaxation to the lower energy states is the 
scattering of two excitons \cite{Tartakovskii}, 
in which one of the final states is a LP. 
We label this scattering process exciton-polariton (XP) scattering.
This relaxation process creates a
non-equilibrium pçolariton distribution that evolves towards a Bose-Einstein
condensate as exciton density is increased\cite{Porras}. Our Hamiltonian 
is the sum of three terms describing the bare-exciton and LP dispersions
($H_0$), the XP scattering ($H_{XP}$), and the
self-interaction in the condensate mode ($H_{SI}$):
\begin{eqnarray}
H_0 = \sum_k \epsilon_k^{LP} a^{\dagger}_k a_k &+&
\sum_k \epsilon_k^{x} b^{\dagger}_k b_k ,
\nonumber \\
H_{XP} = 
\sum_k a^{\dagger}_k F^{\dagger}_k + h.c., &&
H_{SI} =  V a^{\dagger}_0 a^{\dagger}_0 a_0 a_0 .
\label{approxH}
\end{eqnarray} 
The index $k$ is quantized according to $S$. 
$a^{\dagger}_k$ ($\epsilon^{LP}_k$), $b^{\dagger}_k$ ($\epsilon_k^x$) are the
creation operators (energies) of the LP, and bare excitons, respectively. 
$F^{\dagger}_k=
\sum_{k_2,k_3,k_4} 
V_{k,k_2,k_3,k_4}
b^{\dagger}_{k_2} b_{k_3} b_{k_4}$
describes the scattered excitons, 
with $V_{k_1,k_2,k_3,k_4}$ being the polariton-polariton interaction
in the $k a_B << 1$ limit (see \cite{Tassone} for an explicit expression). 
$V = V_{0,0,0,0}$ is the
self-interaction in the ground state.
We have neglected polariton-polariton scattering that involves more than
one lower polariton, and polariton-phonon scattering, because they are much
slower than the process depicted in Fig. \ref{inout}, and only produce energy 
shifts \cite{Porras,Tassone}.

Since we are mainly interested in the evolution of LP, we
trace out the reservoir degrees of freedom in the density matrix
operator $\chi$, and define the reduced density matrix $s$:
\begin{eqnarray}
s(t) = Tr_R \{ \chi(t) \} , \ \ \ 
\langle O_{LP} (t) \rangle = Tr_{LP} \{ O_{LP}(t) s(t)\},
\label{reduceddm}
\end{eqnarray}
where $Tr_R$, $Tr_{LP}$ represent the trace over the exciton reservoir and the
LP, respectively, and $O_{LP}$ is any function of LP operators.

We describe the exciton reservoir by a thermalized Maxwell-Boltzmann
distribution. This approximation is justified by the fast
exciton-exciton scattering within the exciton reservoir. Moreover, this 
assumption is supported by a recent experiment \cite{ring} on  
the evolution of the polariton
distribution when pump-power or temperature is varied.
The total density matrix operator can be approximated by:
\begin{eqnarray}
\chi(t) \approx s(t) \otimes f_R(t) = s(t) \otimes 
e^{\sum_k(\mu_x - \epsilon_k^x) b^{\dagger}_k b_k/
k_B T_x} , 
\label{factordm}
\end{eqnarray}
where $\mu_x$, $T_x$ are the chemical potential and temperature in the exciton
reservoir, respectively.

The time evolution of the density-matrix is calculated 
in the interaction picture. Up to 
the lowest order in $H_{XP}+H_{SI}$, $s$ evolves as:
\begin{eqnarray}
&& \frac{d}{dt} s(t)  = \frac{1}{i \hbar} \left[ H_{SI}, s \right]
- \frac{1}{\hbar^2} \int_{t_0}^t dt' \sum_{k,q} \nonumber \\
&& ( a_k {a^{\dagger}_q}' s' \! - \! {a^{\dagger}_q}' s' a_k ) 
\langle F_k {F^{\dagger}_q}' \rangle \! + \! 
 ( a^{\dagger}_k {a_q}' s' \! - \! {a_q}' s' a^{\dagger}_k )
\langle F^{\dagger}_k {F_q}' \rangle + \nonumber \\
&&
( s'{a^{\dagger}_q}' a_k \! - \! a_k s' {a^{\dagger}_q}' ) 
\langle {F^{\dagger}_q}' F_k  \rangle \! + \! 
( s' {a_q}' a^{\dagger}_k \! - \! a^{\dagger}_k s' {a_q}')
\langle {F_q}' F^{\dagger}_k \rangle ,
\label{chievol2}
\end{eqnarray}
where all operators are in the interaction picture
with respect to $H_{XP} + H_{SI}$, and the primes indicate
time dependence on $t'$.

Eq. (\ref{chievol2}) becomes simpler when LP operators at time $t$ are 
translated to the previous time $t^{\prime}$ by 
$a^{\dagger}_k (t) = a^{\dagger}_k (t') e^{i \epsilon^{LP}_k (t-t')/\hbar}$. 
Moreover, we are interested on the steady-state regime, i. e., in the limit $t_0
\rightarrow - \infty$, so that a Markovian approximation is well justified.
Under this approximation, we take the LP operators in Eq. (\ref{chievol2}) out of the
time integration and obtain the following master equation:
\begin{eqnarray}
\label{lindbladt}
\frac{ds}{dt} &=& 
\frac{i V}{\hbar} \left[ s,{ a_0^{\dagger}}^2 {a_0}^2 \right] + 
\sum_k \left( \frac{W_k^{in}}{2} (a^{\dagger}_k s a_k - a_k a^{\dagger}_k s ) +
\right. \nonumber \\
&& \left. \frac{ W_k^{out}+ \Gamma_k}{2}
(a_k s a^{\dagger}_k - a^{\dagger}_k a_k s )  + h.c. \right) .
\end{eqnarray}
The rates $W^{in}_k$ and $W^{out}_k$ are easily evaluated
with the thermalized exciton distribution \cite{Porras}:
\begin{eqnarray}
W^{in(out)}_k &=&
\sum_{k_2,k_3,k_4} 4 |V_{k,k2,k3,k4}|^2 (1+N^x_{k_2}) N^x_{k_3} N^x_{k_4} 
\nonumber \\
&& \delta (\epsilon_k^{LP} +\epsilon_{k_2}^x
- \epsilon_{k_3}^x - \epsilon_{k_4}^x ) ,
\nonumber \\
W^{out}_k &=&
\sum_{k_2,k_3,k_4} 4 |V_{k,k2,k3,k4}|^2 N^x_{k_2} (1+N^x_{k_3}) (1+N^x_{k_4})
\nonumber \\
&& \delta (\epsilon_k^{LP} +\epsilon_{k_2}^x -
\epsilon_{k_3}^x - \epsilon_{k_4}^x ) ,
\label{Winout}
\end{eqnarray}
where $N^x_k$ are the occupancies in the exciton reservoir.
The imaginary parts of the time integrations have been neglected because they lead to
energy-shifts irrelevant for the process of BEC and phase-diffusion that we consider
here. 
In Eq.(\ref{lindbladt}), we have also included a term (not appearing in 
Eq.(\ref{chievol2})) 
\cite{Scully} that accounts for the radiative
losses with a rate $\Gamma_k = C_k^{LP}/\tau_{ph}$, where $C_k^{LP}$ is the
photon weight in the polariton wave function, and $\tau_{ph}$ is the lifetime of
the photonic mode confined in the microcavity. 

Eq. (\ref{lindbladt}) is the keynote of our analysis. In particular, it 
allows to calculate the evolution (which does not depend on the self-interaction) 
of the LP occupation numbers:  
\begin{eqnarray}
\frac{d}{dt} \langle a^{\dagger}_k a_k \rangle = 
W_k^{in}(\langle a^{\dagger}_k a_k \rangle + 1) - 
(W_k^{out}+\Gamma_k)\langle a^{\dagger}_k a_k \rangle .
\label{Boltzmann}
\end{eqnarray}
We also describe self-consistently the evolution of the parameters 
$n_x$, $T_x$, that describe the exciton reservoir,
by deriving the corresponding rate equations as described in \cite{Porras}.
From the steady  
$n_x$, $T_x$, we get the rates $W_0^{in}$, $W_0^{out}$,
that correspond to the scattering to the ground state. After that, we have found
all the parameters in the master equation (\ref{lindbladt}).

{\it Emission spectrum.}
Once we have a self-consistent description for the time evolution of the density 
matrix, we can undertake the task of computing expectation values of 
magnitudes experimentally measurable. In particular the 
emission spectrum can be obtained from the two-time correlation function:
\begin{equation}
I(\epsilon) = \frac{1}{\pi} Re \int_0^{\infty} 
\langle a^{\dagger}_0 (\tau) a_0(0) \rangle 
e^{-i \epsilon \tau / \hbar} d \tau .
\label{spectrum}
\end{equation}
Without self-interaction in Eq. (\ref{lindbladt}),
the application of the quantum regression theorem would lead trivially to 
Lorentzian line shape of the spectrum:
\begin{eqnarray}
\langle a^{\dagger}_0 (\tau) a_0(0) \rangle^{NI} \! = \! N_0
e^{-\Gamma^{NI} \tau } , \ 
\Gamma^{NI} \! = \! \frac{W_0^{out}+\Gamma_0}{2(1+N_0)} .
\label{exponential}
\end{eqnarray}
Decoherence has the usual aspect of a single-particle noise corrected, 
in the denominator, by the number
of particles in the condensate. However, inclusion of the self-interaction term
changes this result dramatically. We use a well-known method in quantum optics 
that allows to include exactly the effect of the self-interaction\cite{Scully}.  
The two-time average is expanded as a sum over a set of auxiliary 
functions $C_n(\tau)$:
\begin{eqnarray}
&& \langle a^{\dagger}_0(\tau) a_0(0) \rangle
\! = \! \sum_n \sqrt{n} C_n(\tau) ,
\nonumber \\
&& C_n(\tau) \! = \! \langle e^{iH_0 \tau /\hbar} \! |n \! > \! < \! n-1 \! | 
e^{-iH_0 \tau/ \hbar } a_0(0) \rangle ,
\label{correlation}
\end{eqnarray} 
with $|n>$ being the number representation of the $k=0$ mode. 
One can use the quantum regression theorem, to show that the functions 
$C_n$ satisfy the differential equation \cite{Holland}:
\begin{equation}
\frac{d}{d \tau} C_n(\tau) = \frac{i}{\hbar} \epsilon^{LP}_0 C_n(\tau) + 
\sum_m L_{n,m} C_m (\tau) ,
\label{Cevol}
\end{equation}
where the $L_{n,m}$ are the coefficients governing the evolution of 
off-diagonal one-time matrix elements:
\begin{equation}
\frac{d}{d \tau} \! < n-1| s(\tau) | n > = \! \sum_{m} L_{n,m} 
\! < m-1 | s(\tau) | m> .
\label{Qevol}
\end{equation} 
Using Eqs.(\ref{lindbladt}) and (\ref{Qevol}), Eq. (\ref{Cevol}) becomes:
\begin{eqnarray}
&&\frac{d}{d \tau} C_n(\tau) =
\left( - w^+_0 \ (2n+1) \right. - w^-_0 \ (2n-1) +
\nonumber \\ 
&&\frac{i}{\hbar}V(n-1-N_0)\left. \right) C_n(\tau) + 
w^+_0 \ 2 \sqrt{n(n-1)} C_{n-1}(\tau)
\nonumber \\
&& + w^-_0 \ 2 \sqrt{n(n+1)} C_{n+1}(\tau) ,
\label{discrete}
\end{eqnarray}
with $w^+_0=W^{in}_0/2$ and $w^-_0=(W^{out}_0+\Gamma_0)/2$.
In Eq. (\ref{discrete}) we have taken the origin of energies at 
$\epsilon^{LP}_0+V N_0$, in order to
compare the linewidths of the emission spectrum at different densities.
Initial condition for Eq. (\ref{discrete}) is obtained from 
$C_n$ for $\tau =0$:
\begin{equation}
C_n(0) = \sqrt{n} < n | s | n > =
\sqrt{n} \left( 1 \! - \! \frac{w_0^+}{w_0^-} \right)
\left( \frac{w_0^+}{w_0^-} \right)^n \! ,
\label{initial}
\end{equation}
where $< n | s | n >$ is easily deduced from having a time-derivative equal to 
zero in the steady state.
Instead of solving numerically the enormous set of Eqs. (\ref{discrete}), 
the problem can be simplified,
in the case $N_0 >> 1$,
by replacing the index $n$ by a continuous variable
\cite{Gardiner}, obtaining a partial differential equation:
\begin{eqnarray}
\label{continuous}
&&\frac{\partial C(n,\tau)}{\partial \tau} =
n(w^-_0 + w^+_0) \frac{\partial^2}{\partial n^2} C(n,\tau) +
\nonumber \\
&&\left( 2 n ( w^-_0 - w^+_0 ) + w^-_0 + w^+_0 \right) 
\frac{\partial}{\partial n} C(n,\tau) + \\
&&\left( 2(w^-_0 - w^+_0) - \frac{w^-_0+w^+_0}{4n} + 
\frac{iV}{\hbar}(n-N_0-1) \right) C(n,\tau) . 
\nonumber 
\end{eqnarray}
In the limit $\hbar \Gamma^{NI}>>V N_0$, 
(\ref{continuous}) gives the adequate Lorentzian shape of the spectrum with a 
linewidth $\Gamma^{NI}$. In the opposite limit, $V N_0 >> \hbar \Gamma^{NI}$,
an analytic solution exists:
\begin{equation}
C(n,\tau) \approx \frac{1}{N_0} \sqrt{n} e^{-\frac{n}{N_0} 
+ i\frac{V}{\hbar}(n-N_0)\tau} ,
\label{approx}
\end{equation}
predicting an asymmetrical lineshape:
\begin{equation}
I(\epsilon) \approx \frac{\hbar}{V} 
\frac{(\epsilon + V N_0)}{VN_0}
e^{-(\epsilon + V N_0)/(V N_0)} \theta(\epsilon + V N_0) .
\label{approxspectrum}
\end{equation}

\begin{figure}
\centerline{
\psfig{figure=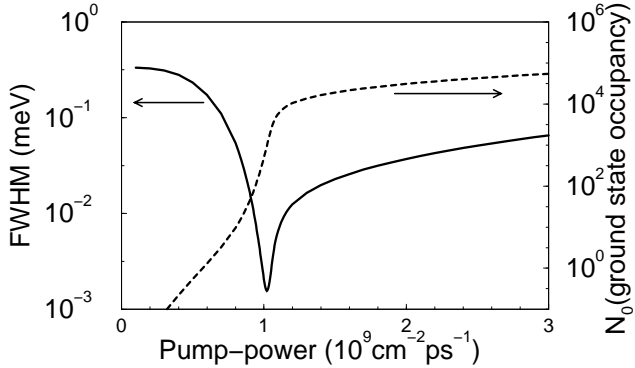,height=2.0in,width=3.0in}
}
\caption{Full Width at Half Maximum of the emission spectra (continuous line) 
and ground state occupancy (dashed line) as a function of the pump-power $p_x$ 
(in $10^9 cm^{-2} ps^{-1}$). 
The optimum pump corresponds to $p_x \approx 10^9 cm^{-2} ps^{-1}$.}
\label{linewidth}
\end{figure}

{\it Results for a CdTe microcavity.}
We have solved numerically Eq. (\ref{continuous}) for the case of a CdTe
microcavity with $\Omega_P=10$meV, and zero detuning between
the bare exciton and the photonic mode. 
The pump is assumed to add excitons at a 
a given rate, $p_x$, and at a lattice temperature, $T_L=10K$. 
This implies a very fast relaxation by the exciton-phonon scattering,
which is not always
the case in experiments. The steady value of $T_x$, however can reach 30 K, as
explained in \cite{Porras}.
Lifetimes of the photon and the bare excitons are taken $\tau_{ph}=1ps$ and
$\tau_x=100ps$. The steady-state polariton density considered in our calculation
is always below $0.3$ times the saturation density, which can be estimated to be
$6.7 \times 10^{11} cm^{-2}$ in a CdTe microcavity \cite{Porras}.
The quantization length is
$50 \mu m$, of the order of typica l excitation spot diameters.
Fig. \ref{linewidth} gives  $N_0$ as a function of the pump-power. It 
shows a threshold for BEC ($N_0>1$) around $p_x  \approx 8 10^8 cm^{-2}ps^{-1}$.

\begin{figure}
\centerline{
\psfig{figure=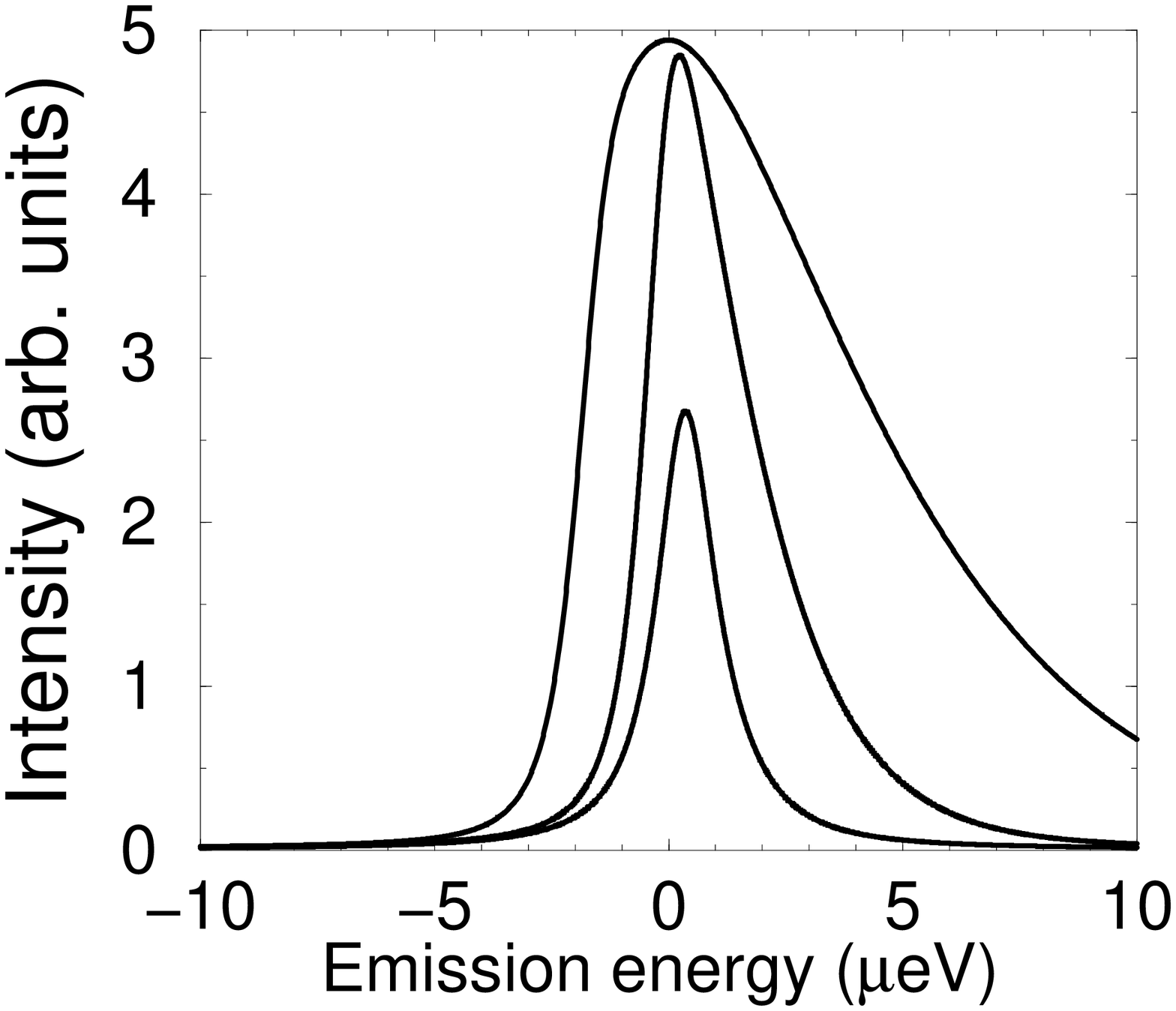,height=1.5in,width=1.5in}
\psfig{figure=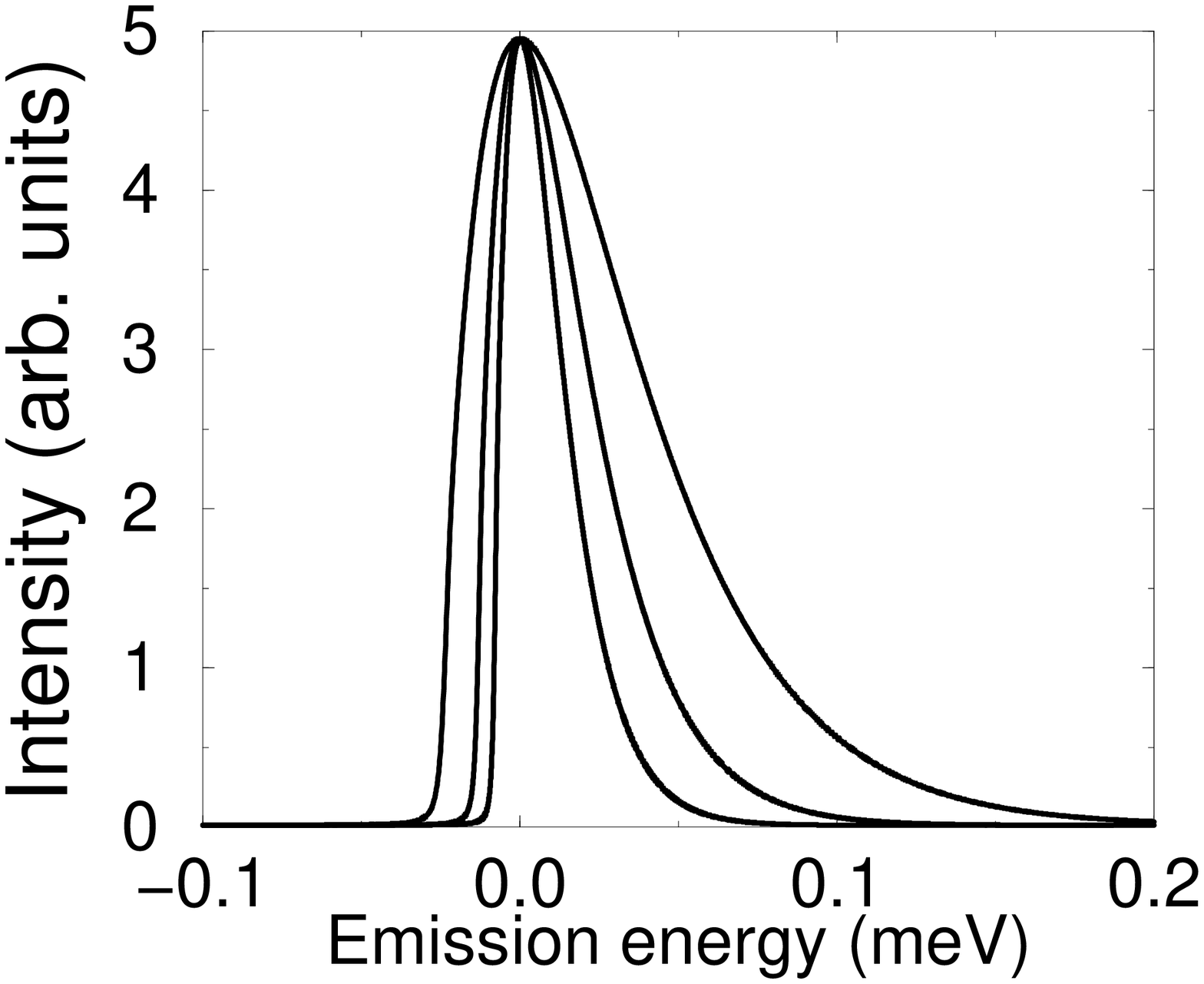,height=1.5in,width=1.5in}
}
\caption{Left: Transition from non-interacting, to self-interaction 
decoherence in the emission spectra for $p_x = 1.02 (inner), 1.05, 1.10 (outer)$
in units 
of $ 10^9 cm^{-2}ps^{-1}$.  Right:
Continuous increase of the linewidth shown in the spectra for
$p_x = 1.5 (inner), 2.0, 3.0 (outer)$ in units of $ 10^9 cm^{-2}ps^{-1}$.}
\label{spectra}
\end{figure}

For densities below the condition $N_0 V \approx \hbar \Gamma^{NI}$, the 
calculated lineshape is Lorentzian with minimum linewidth 
of the order of $1 \mu eV$. After condition $N_0 V \approx \hbar \Gamma^{NI}$ 
is reached, there is a transition from the laser-like decoherence to the 
self-interaction broadening as shown in Fig. \ref{spectra} (left). For larger
pump-powers, the emission is asymmetric and is continuously broadened. Above 
$p_x = 1.1 10^{9} cm^{-2}ps^{-1}$, the numerical results are identical to the
approximation given in Eq. (\ref{approxspectrum}). 
In the evolution of the spectrum linewidth as a function of the pump-power, 
shown in Fig. \ref{linewidth}, one observes that the optimum pump is very well 
defined by the abrupt dip in the emission linewidth. This dip means an increase 
of the coherence of almost three orders of magnitude. 
For larger densities the linewidth of the polariton laser increases
linearly with the number of condensed particles, until it reaches values
comparable to the non-condensed emission.

In conclusion, we have presented a theoretical method for the self-consistent 
calculation of the emission linewidth of a polariton BEC.
Self-interaction sets an important restriction for the
coherence that can be achieved in this system. 
The dynamics described by our Eq. (\ref{lindbladt}) would drive the system to a BEC
provided a small coherent seed, i.e. a symmetry-breaking term, is included in the
initial condition.
Our conclusions are
also relevant for the case of recent proposals in which the scattering
mechanism for the relaxation of polaritons towards the ground state is different
than the XP scattering considered here, as for instance the case of the proposed 
electron-polariton scattering in doped microcavities \cite{Malpuech}.

We thank H. Kohler and J. Fern\'andez-Rossier 
for fruitful discussions.
Work supported in part by MCYT of Spain under contract
MAT2002-00139 and CAM under contract 07N/0042/2002.

\widetext


\begin{references}

\bibitem{parametric} P.G. Savvidis {\it et al.}, Phys. Rev. Lett. {\bf 84}, 
1547 (2000). J.J. Baumberg {\it et al.}, Phys. Rev. B {\bf 62}, R16247 (2000).

\bibitem{Ciuti} C. Ciuti {\it et al.}, Phys. Rev. B {\bf 62}, R4825 (2000). 
C. Ciuti, P. Schwendimann, and A. Quattropani, Phys. Rev. B {\bf 63}, 041303 (2001).

\bibitem{Imamoglu} A. Imamoglu {\it et al.},
Phys. Rev. A {\bf 53}, 4250 (1996).

\bibitem{Eastham} P.R. Eastham and P.B. Littlewood, Phys. Rev. B. {\bf 64}, 235101
(2001).

\bibitem{Deng} H. Deng {\it et al.}, Science {\bf 298}, 199 (2002).

\bibitem{CdTe} Le Si Dang {\it et al.} Phys. Rev. Lett. {\bf 81}, 3920 (1998).
M. Saba {\it et al.}, Nature (London) {\bf 414}, 731 (2002), 
M.D. Mart\'\i n {\it et al.}, Phys. Rev. Lett. {\bf 89}, 077402 (2002).

\bibitem{Zamfirescu} M. Zamfirescu {\it et al.}, Phys. Rev. B {\bf 65}, 161205
(2002). 

\bibitem{Porras} D. Porras {\it et al.}, Phys. Rev. B {\bf 66}, 085304 (2002).

\bibitem{Scully} M.O. Scully and M.S. Zubairy, 
{\it Quantum Optics}, Cambridge University Press, Cambridge (1997).

\bibitem{Holland} M. Holland {\it et al.}, Phys. Rev. A {\bf 54}, R1757 (1996).

\bibitem{GardinerIII} C.W.Gardiner {\it et al.}, Phys. Rev. A {\bf 58}, 536
(1998).  

\bibitem{Tassone00} F. Tassone and Y. Yamamoto, Phys. Rev. A {\bf 62}, 063809
(2000).

\bibitem{Malpuech2} A. Kavokin {\it et al.} Phys. Lett. A {\bf 306}, 187 (2003).

\bibitem{Hopfield} J.J. Hopfield, Phys. Rev. {\bf 112}, 1555 (1958).

\bibitem{Tartakovskii} A.I. Tartakovskii {\it et al.}, 
Phys.Rev. B {\bf 62}, R2283 (2000).

\bibitem{Tassone} F. Tassone and Y. Yamamoto, Phys. Rev. B {\bf 59}, 10830 (1999). 

\bibitem{ring} P.G. Savvidis {\it et al.}, Phys. Rev. B {\bf 65}, 073309 (2002).

\bibitem{Gardiner} C.W. Gardiner, {\it Handbook of Stochastic Methods}, Springer,
Berlin (1996).

\bibitem{Malpuech} G. Malpuech {\it et al.}, Phys. Rev. B {\bf 65}, 153310 (2002). 



\end{references}
\end{document}